\documentstyle[12pt]{article}

\ifx\text\undefined
\let\text\textrm    
\fi

\begin{document}

\title{Anomalous Viscosity, Resistivity, and Thermal Diffusivity of the Solar
Wind
Plasma}
\author{Mahendra K. Verma  \\ 
Department of Physics, Indian Institute of Technology, Kanpur 208016, India}
\date{\today }
\maketitle

\begin{abstract}
In this paper we have estimated typical anomalous viscosity, resistivity,
and thermal difffusivity of the solar wind plasma. Since the solar wind is
collsionless plasma, we have assumed that the dissipation in the solar wind
occurs at proton gyro radius through wave-particle interactions. Using this
dissipation length-scale and the dissipation rates calculated using MHD
turbulence phenomenology [{\it Verma et al.}, 1995a], we estimate the
viscosity and proton thermal diffusivity. The resistivity and electron's
thermal diffusivity have also been estimated. We find that all our transport
quantities are several orders of magnitude higher than those calculated
earlier using classical transport theories of {\it Braginskii}. In this
paper we have also estimated the eddy turbulent viscosity.
\end{abstract}

\pagebreak

\section{Introduction}

The solar wind is a collisionless plasma; the distance travelled by protons
between two consecutive Coulomb collisions is approximately 3 AU [{\it %
Barnes, }1979]. Therefore, the dissipation in the solar wind involves
wave-particle interactions rather than particle-particle collisions. For the
observational evidence of the wave-particle interactions in the solar wind
refer to the review articles by {\it Gurnett }[1991], {\it Marsch }[1991]
and references therein. Due to these reasons for the calculations of
transport coefficients in the solar wind, the scales of wave-particle
interactions appear more appropriate than those of particle-particle
interactions [{\it Braginskii}, 1965].

Note that the viscosity in a turbulent fluid is scale dependent. The
viscosity discussed in most part of this paper is the one at the dissipation
length-scale, not at the large or intermediate length-scales. The viscosity
at large scale is called turbulent eddy viscosity; it is briefly discussed
at the end of section 2 of this paper. In fluid turbulence, the viscosity at
dissipation scales is determined from the dissipation rate and the
dissipation length scale [{\it Lesieur, }1989]. In this paper we estimate
viscosity, resistivity, and thermal conductivity of the solar wind using
similar technique. The dissipation length-scales is obtained from the
wave-particle interactions, and the dissipation rates are obtained from the
Kolmogorov-like MHD turbulence phenomenology [{\it Marsch, }1990; {\it %
Matthaeus and Zhou, }1989; {\it Zhou and Matthaeus, }1990]. For the solar
wind {\it Tu }[1988] and {\it Verma et al. }[1995a] have calculated the
dissipation rates from the observed energy spectra using the Kolmogorov-like
MHD turbulence phenomenology. Since in our approach the wave-particle
interactions dominate the particle-particle collisions, the transport
coefficients presented here are the anomalous transport coefficients
commonly referred to in fusion plasma literature. Note that the transport
quantities in the solar wind vary with distance. In this paper we estimate
these quantities at 1\ AU.

Earlier {\it Montgomery} [1983] has calculated the transport coefficients in
the solar wind using the {\it Braginskii}'s [1965] formalism, which is based
on particle-particle collision. He found both kinematic viscosity and
resistivity to be of the order of $10^{-6}$ km$^2$ s$^{-1}$. Using the
velocity of the large eddies as 20 km s$^{-1}$ and length-scale as 10$^6$
km, he obtained the Reynolds number to be of the order of $10^{13}$. Note
that {\it Montgomery }[1983] used $\eta _1$ of {\it Braginskii} rather than $%
\eta _0$. This is consistent with the {\it Hollweg'}s{\it \ }result [1985]
where he showed that the $\eta _0$ terms are fully accounted for by the
diagonal pressure tensor.

In the following section we will estimate the length-scale at which
wave-particle interactions take place. This will be the dissipation
length-scale for our calculation of the transport coefficients. Using the
dissipation rates calculated earlier, we then estimate the kinematic
viscosity, resistivity, and thermal diffusivity at dissipation
length-scales. Towards the end of section 2, we have also estimated the eddy
viscosity and thermal diffusivity of the solar wind. Section 3 contains
conclusions.

\section{Calculation}

{\it Verma et al.} [1995a] have calculated the dissipation rates in the
solar wind streams using the Kolmogorov-like MHD turbulence phenomenology [%
{\it Marsch, }1990; {\it Matthaeus and Zhou, }1989; {\it Zhou and Matthaeus,
}1990]. The choice of this phenomenology over the Kraichnan's phenomenology [%
{\it Kraichnan,} 1965] or the Dobrowolny et al.'s generalization of
Kraichnan's phenomenology [{\it Dobrowolny et al., }1980] is motivated by
the fact that the observed solar wind energy spectra tend to be closer to
Kolmogorov's $k^{-5/3}$ power law than Kraichnan's $k^{-3/2}$ power law.
Also, temperature evolution study of {\it Verma et al. }[1995a] show that
the predictions of the temperature evolution using the Kolmogorov-like model
are in closer agreement with the observations than those using Kraichnan's
or Dobrowolny et al.'s models. Also refer to {\it Tu }[1988] for theoretical
studies of turbulent heating in the solar wind.

The Kolmogorov-like phenomenology provides the energy spectra of
fluctuations ${\bf z}^{\pm }={\bf u\pm b/}\sqrt{4\pi \rho }$, where ${\bf u}$
is the velocity field fluctuation, ${\bf b}$ is the magnetic field
fluctuation, and $\rho $ is the density of the plasma. The quantities ${\bf z%
}^{\pm }$ represent the amplitudes of Alfv\'en waves having positive and
negative velocity-magnetic field correlations respectively. The energy
spectra according to this phenomenology are
\begin{equation}
\label{kolmlike}E^{\pm }(k)=K^{\pm }\left( \epsilon ^{\pm }\right)
^{4/3}\left( \epsilon ^{\mp }\right) ^{-2/3}k^{-5/3}.
\end{equation}
where $\epsilon ^{\pm }$ are the dissipation rates of ${\bf z}^{\pm }$
fluctuations, and $K^{\pm }$ are Kolmogorov's constants for MHD turbulence.
According to {\it Verma et al. }[1995a] the dissipation rates of the solar
wind streams are of the order of $10^{-3}$ km$^2$ s$^{-3}$.

As mentioned in the introduction, we estimate the dissipation length-scale
from the theories of wave-particle interactions. It has been shown that the
wave-particle resonance between MHD waves and ions occurs either in the form
of the Doppler-shifted cyclotron resonance,
\begin{equation}
\label{cyclo}\omega -k_{\parallel }v_{\parallel }=n\Omega _i;\left( n=\pm
1,\pm 2,\cdot \cdot \cdot \right)
\end{equation}
or in the form of the Landau resonance
\begin{equation}
\omega -k_{\parallel }v_{\parallel }=0,
\end{equation}
where $\Omega _i$ is the cyclotron frequency of the ions, $\omega $ is the
wave frequency, $k_{\parallel }$ and $v_{\parallel }$ are the parallel
components along the mean magnetic field of the wave number and the ion
velocity vector respectively [{\it Stix, }1962; {\it Barnes, }1979]. For the
solar wind, at 1 AU the cyclotron frequency $\Omega _i$ of the ions is of
the order of 1.0 s$^{-1}$, and the thermal speed $v$ is of the order of 50
km s$^{-1}$ [{\it Barnes, }1979]. Typical Alfv\'en speed $\omega /k$ at 1\
AU is also 50 km s$^{-1}$. Also note that the solar wind fluctuations are
dominated by Alfv\'en waves; the compressive waves are damped at the early
stages of its transit.

The process by which Alfv\'en waves might be damped have been the subject of
considerable research. Since $\omega /k\sim 50$ km/s $\sim v$, it appears
that Alfv\'en waves can be Landau damped. However, {\it Barnes and Suffolk }%
[1971] and {\it Barnes }[1979] argue against this. They show that the
transverse Alfv\'en waves are exact solutions of the Vlasov-Maxwell
equations for arbitrary amplitude, hence it can not be damped. But that is
not correct either. It has been shown that all hydromagnetic waves, except
the Alfv\'en mode with precise circular polarization, steepen and evolve
into other modes or collisionless shocks [{\it Tidman and Krall}, 1971].
{\it Sagdeev and Galeev }[1969] showed that a linearly polarized Alfv\'en
wave is unstable and it decays to a back scattered Alfv\'en wave and
magnetosonic waves. The magnetosonic waves thus generated get damped by
Landau damping (see {\it Barnes }[1979] and references therein for
discussion on Landau damping of magnetosonic waves). {\it \ Hollweg }[1971]
has obtained similar results. Hence the Alfv\'en waves in the solar wind can
get damped by decaying to a magnetosonic waves which in turn gets damped by
Landau damping.

Now the question arises, which waves in the solar wind are affected by the
above process. The energy from the small and intermediate $k$ (large
wavelength) waves cascades to larger $k$ waves due to nonlinear interaction
arising from the ${\bf z}^{\mp }\cdot \nabla {\bf z}^{\pm }$ term of MHD
equation [{\it Kraichnan, }1965], and these waves do not get damped. At the
dissipation scale the energy cascade stops. We conjecture that the decay of
Alfv\'en waves to magnetosonic waves, and the damping of the generated
magnetosonic waves occur near the ion gyro radius $r=100$ km. Therefore, $%
k_d=10^{-2}$ km$^{-1}$.

Regarding the cyclotron resonance, the small $k$ Alfv\'en waves of the solar
wind cannot be damped by this mechanism because $w\ll \Omega _i$ and $kv\ll
\Omega _i$, when $k$ is small [see Eq. (\ref{cyclo})]. However, when $k$
becomes large, it is possible for the waves to get damped by cyclotron
damping. The approximate $k$ where cyclotron resonance could happen is
\begin{equation}
k_d\sim \frac{\Omega _i}{V_A-V_{\parallel }}\sim \frac 1{50km}\sim
10^{-2}km^{-1}.
\end{equation}
Hence the dissipation length-scale for the particle-wave interaction is
approximately 100 km, and the dissipation wavenumber is $k_d=10^{-2}$ km$%
^{-1}$. The solar wind observations show that at 1 AU the transition from
inertial range to dissipation range occurs at around a length scale of 400
kms [{\it Roberts, }1995], a result consistent with our above arguments. In
this paper we assume that the dissipation length-scales for fluid energy,
magnetic energy, and the energy of the Alfv\'en waves are all same (see
Appendix).

In the appendix we derive an expression for the viscosity $\nu $ in terms of
dissipation rate $\epsilon $ and dissipation length-scale $(k_d^{-1})$ that
is
\begin{equation}
\nu \sim \left( \frac \epsilon {k_d^4}\right) ^{1/3}.
\end{equation}
Here we assumed that the fluid and magnetic energies are approximately
equal, and also that $E^{+}(k)\sim E^{-}(k)$. Under this condition $\nu \sim
\lambda .$ We use this formula for our estimation of viscosity in the solar
wind. Substitution of $\epsilon =10^{-3}$ km$^2$s$^{-3}$ and $k_d=10^{-2}$ km%
$^{-1}$ in the above equation yields $\nu \sim 50$ km$^2$ s$^{-1}$. This
result is very different from the one obtained by {\it Montgomery} [1983].
Note that  the above viscosity is the ion viscosity. It is interesting to
note that our estimate of ion viscosity is close to Bohm's diffusion
coefficient [{\em Chen, }1974], which is
\begin{equation}
D_B=\frac{k_BTc}{16eB}\sim 100\ km^2s^{-1}
\end{equation}
where $k_B$ is the Boltzmann constant, $T$ is the proton temperature, $c$ is
the speed of light, $e$ is the electronic charge, and $B$ is the mean
magnetic field. The Reynolds number with $\nu =50$ km$^2$s$^{-1}$, the mean
speed $U=20$ km s$^{-1}$, and the length-scale of 10$^7$ km is
\begin{equation}
Re=\frac{UL}\nu =4\times 10^6.
\end{equation}
The dissipation time-scale is
\begin{equation}
\tau _d\sim \frac 1{k_dv_d}\sim \frac 1{\left( k_d^2\varepsilon \right)
^{1/3}}\sim 200\ \text{s}
\end{equation}
where $v_d$ is the velocity at the dissipation scale. For the above
expression, we assumed that the Kolmogorov-like MHD turbulence phenomenology
(Eq. (\ref{kolmlike})) is valid till $k=k_d$, therefore, $v_d\sim
(E(k_d)k_d)^{1/2}\sim (k_d/\epsilon )^{1/3}$ [{\it Lesieur}, 1989].

Now we estimate electron viscosity and resistivity. The classical electron
viscosity $\nu _e$ will be [{\it Braginskii}, 1965]
\begin{equation}
\nu _e=\left( \frac{m_e}{m_i}\right) ^{1/2}\nu _i^1\sim 2\times
10^{-8}km^2s^{-1}.
\end{equation}
{\it Scudder and Olbert }[1979a, b] have shown that classical collision
transport applies partly to the solar wind `core' electrons, but are
inappropriate for the `halo' electrons which are affected by whistler waves.
However, {\it Schwartz et al.} [1981] showed that waves with frequencies
near the ion gyrofrequencies and wave vectors comparable with inverse ion
Larmor radii can provide strong electron-wave coupling. Since at this moment
the results are not conclusive, we are following {\it Schwartz et al.}'s
[1981] paradigm.

In the following discussion we will attempt to define and estimate
resistivity and electron viscosity in a turbulent plasma. The arguments
presented here is over simplified and speculative, and they are in similar
lines as that of {\it Priest }[1982]. However, we believe that these
arguments shed some light into this complicated problem and will be useful
for future development in this area. The arguments follow: Eddy or kinematic
viscosity can be interpreted as diffusion coefficient for coherent fluid
parcels. The dissipation length-scale discussed in this paper is the
length-scale where the smallest coherent fluid parcel disperse, i.e., fluid
energy after this scale is zero (refer to Appendix). Similarly, the coherent
magnetic structures are destroyed by resistivity at the dissipation scales.
Since the resistivity is dominated by electron's transport properties,  here
we estimate electron resistivity. Also since the solar wind plasma is
turbulent, we use the length-scales of interactions of the electrons with
the waves as the relevant length-scale for this purpose. We assume in this
paper that the dissipation length-scales of both fluid and magnetic energy
are $k_d^{-1}$. Therefore, $l_d^e=$ $k_d^{-1}\sim 100$ km. Since the
electrons are lighter particles, they move much faster than protons; we
assume that the relevant speed of the electrons at dissipation length-scale
is its thermal speed. Taking electron temperature as $10^5$ K, $v_d^e=1000$
km/s. From these two scales, we can obtain the time-scale that is $\tau
_d^e\sim l_d^e/v_d^e\sim 0.1$ sec. Our above arguments are in the same
spirit as that of {\it Priest }[1982] who has obtained a formula for
anomalous conductivity using anomalous collision-time.

Using the above electron velocity and length scales we obtain the electron's
kinematic viscosity that is $\nu \sim v_d^el_d^e\sim 10^5$ km$^2$s$^{-1}$.
We can also estimate the resistivity using the above estimates of length and
time scales. The resistivity $\lambda $ is defined as [{\it Braginskii, }%
1965]
\begin{equation}
\lambda =\frac{m_ec^2}{4\pi ne^2\tau _d^e}
\end{equation}
where $m_e$ is the mass of the electron. Substitution of $n=5$ ions/cc and $%
\tau _d^e=0.1$ sec in the above expression yields $\lambda \sim 100$ km$^2$s$%
^{-1}$. The resistivity calculated here is close to the resistivity
calculated in the earlier part of the paper using the dissipation rates,
hence our calculations appear consistent. The magnetic Reynolds number will
be
\begin{equation}
Re_M=\frac{UL}\lambda \sim 2\times 10^6.
\end{equation}
The solar wind magnetic Prandtl number, defined as $\nu /\lambda $, appears
to be of the order of unity. It is interesting to note that both
renormalized viscosity $\nu (k)$ and resistivity $\lambda (k)$ are expected
to scale as $\epsilon ^{1/3}k^{2/3}$, where $\epsilon $ is the relevant
dissipation rate, and $k$ is the wavenumber [{\it Verma and Bhattacharjee, }%
1995b and references therein]. Therefore, renormalized magnetic Prandtl
number $\nu (k)/\lambda (k)$ $\sim 1.$ In this paper we are calculating $%
\lambda (k_d)\ $and $\nu (k_d).$ It is reasonable to expect that
Kolmogorov's 5/3 power law continues till $k=k_d$, therefore, it is not
surprising that our magnetic Prandtl number $\lambda (k_d)/\nu (k_d)$ $\sim
1 $. However, since the above numbers are only order of magnitude estimates,
we can not make definite prediction about the magnetic Prandtl number.

The anomalous thermal diffusivity $\kappa _{i,e}$ of ions and electrons of
the heat diffusion equation [{\it Priest, }1982; {\it Landau, }1987]
\begin{equation}
\frac{\partial T_{i,e}}{\partial t}=\kappa _{i,e}\nabla ^2T_{i,e}
\end{equation}
can also be estimated from the above dissipation length and time scales. It
has been argued earlier that heat conduction is carried by superthermal
electrons [{\it Marsch, }1991 and references therein]. Also, observational
studies by {\it Philip et al.} [1987] show that the heat flux is not
proportional to the electron temperature gradient but are regulated by
whistler-mode instability driven by the skewness of the distribution
function [{\it Gary et al.}, 1994]. However, here we use turbulence scaling
arguments because of  the reasons stated above. Since some of the issues are
not settled yet, e.g., the heat flux calculated by classical transport
quantities are not in good agreement with the observed heat flux, we
estimate turbulent thermal diffusivity to provide another point of view.

Here again we use wave-particle interaction time-scale rather the
particle-particle collision time-scales. By dimensional arguments the
coefficient $\kappa _{i,e}$ can be approximated by
\begin{equation}
\kappa _{i,e}\sim \frac 1{k_d^2\tau _{i,e}}
\end{equation}
The substitution $k_d$ and $\tau _{i,e}$ of the solar wind in the above
equation yields $\kappa _i\sim $ 50 km$^2$ s$^{-1}$ and $\kappa _e\sim $ $%
10^5$ km$^2$ s$^{-1}$. These numbers are same as the viscosity calculated
above. The ratio $\nu /\kappa $ is called Prandtl number, and it is of the
order unity. Similar to viscosity, the thermal diffusivity calculated here
are orders of magnitude higher than the one calculated from the Braginskii's
formalism in which $\kappa _i\sim $ 10$^{-6}$ km$^2$ s$^{-1}$ and $\kappa
_e=\kappa _i(m_e/m_i)^{1/2}\sim $ 2$\times $10$^{-8}$ km$^2$ s$^{-1}$(same
as the viscosity of {\it Montgomery }[1983]).

As mentioned in the introduction, viscosity is scale-dependent. The
large-scale viscosity, called eddy viscosity, is $\nu _L\sim v_LL$, where $L$
is the large-scale length and $v_L$ is the large-scale fluctuating speed.
Therefore, for the solar wind $\nu _L\sim 20km/s\times 10^8km\sim 10^9km^2/s$%
. This number is seven orders of magnitude higher than the viscosity at
dissipation length-scale. The thermal diffusivity at large-scales is
approximately equal to the eddy viscosity. These quantities could be useful
for the study of solar wind evolution of energy and heat flux etc.

\section{Conclusions and Discussion}

In this paper we have calculated the viscosity, resistivity, and thermal
diffusivity of the solar wind using nonclassical approach. The solar wind is
collisionless, therefore, the wave-particle interactions become important
while considering dissipation mechanisms in the wind. In this paper we have
fixed the dissipation length-scale at proton gyro radius ($\sim 100$ kms),
scale at which wave-particle interactions are expected to occur. This result
is consistent with the solar wind observations in which the transitions from
inertial range to dissipation range at 1 AU occur at around 400 kms. In our
calculation we also need turbulent dissipation rates occurring in the solar
wind. In this paper we take the turbulent dissipation rates calculated by
{\it Verma et al. }[1995a].

We find that a typical ion viscosity is 50 km$^2$ s$^{-1}$ and electron
viscosity is 2$\times 10^5$ km$^2$ s$^{-1}$. The corresponding Reynolds
number (with ion viscosity) is around 10$^6$. The resistivity is around 200
km$^2$ s$^{-1}$, and the magnetic Reynolds number is also around 10$^6$. The
magnetic Prandtl number is order unity. The ion and electron thermal
diffusivities are the same as the ions and electron viscosities
respectively. The large-scale (eddy) viscosity of the wind is approximately
10$^9$ km$^2/$s.

All the transport quantities calculated by us are several orders of
magnitude higher than those calculated earlier using classical transport
theory of {\it Braginskii. }These results should have important consequences
on modelling of the solar wind. Our results show that the thermal
diffusivity in the solar wind is much higher than what have been assumed
earlier and should be important for the studies regarding the temperature
evolution of the solar wind.

The author thanks J. K. Bhattacharjee, D. A. Roberts, M. L. Goldstein, and
J. F. Drake for discussion and comments.

\appendix{}

In this appendix we derive an expression for viscosity in terms of energy
dissipation rates and dissipation length-scales. We use energy equation to
derive this expression. Incompressible MHD equation in absence of a mean
magnetic field is [{\it Kraichnan}, 1965]
\begin{equation}
\frac \partial {\partial t}{\bf z}^{\pm }=-{\bf z}^{\mp }\cdot \nabla {\bf z}%
^{\pm }-\nabla p+\nu _{+}\nabla ^2{\bf z}^{\pm }+\nu _{-}\nabla ^2{\bf z}%
^{\mp }
\end{equation}
\begin{equation}
{\bf z}^{\pm }={\bf u\pm b}
\end{equation}
\begin{equation}
\nu _{\pm }=\frac 12\left( \nu \pm \lambda \right)
\end{equation}
where {\bf u} is the fluctuating velocity field, {\bf b }is the fluctuating
magnetic field in velocity units, $p$ is the total pressure, $\nu $ is the
kinematic viscosity, and $\lambda $ is the resistivity. From this equation,
under the assumption of isotropy of fluctuations, one can derive [{\it %
Orszag,} 1977]
\begin{equation}
\frac \partial {\partial t}E^{\pm }(k)=-2\nu _{+}k^2E^{\pm }(k)-2\nu
_{-}k^2(E^u(k)-E^b(k))+T^{\pm }(k)
\end{equation}
where $E^{\pm }(k)$ are the energy spectra of ${\bf z}^{\pm }$, $E^u(k)$ and
$E^b(k)$ are the velocity and magnetic field energy spectra respectively,
and $T^{\pm }(k)$ comes from nonlinear term and involves triple correlations
of ${\bf z}^{\pm }$. By integrating the above equation over the whole
spectrum, we obtain
\begin{equation}
\label{energy}\epsilon ^{\pm }=-2\nu _{+}\int_0^\infty k^2E^{\pm }(k)dk-2\nu
_{-}\int_0^\infty k^2(E^u(k)-E^b(k))dk.
\end{equation}
The term $T^{\pm }(k)$ upon integration over the whole spectrum yields zero [%
{\it Orszag,} 1977].

We make several assumptions to get an order of magnitude estimates of $\nu $%
. We assume that the third term of the above equation vanishes. This
condition will be satisfied either if $\nu _{-}=0$ or $E^u(k)=E^b(k)$. Since
the spectra $E^{\pm }(k)$ is usually strongly damped in the dissipation
range, most contribution to the first integral of the above equation comes
from $k$ in the range of $0$ to $k_d$. We also make a drastic assumption
that $E^{+}(k)=E^{-\;}(k)$, and $\epsilon ^{+}=\epsilon ^{-}=\epsilon $.
These assumptions are justified only because we are making order of
magnitude estimation of $\nu $. To obtain somewhat precise values of $\nu
_{\pm },$ we will have to analyse Eq. (\ref{energy}) carefully. Now the
substitution Eq. (\ref{kolmlike}) for $E^{\pm }(k)$ in Eq. (\ref{energy})
yields
\begin{equation}
\nu \sim \lambda \sim \nu _{+}\sim \left( \frac \epsilon {k_d^4}\right)
^{1/3}
\end{equation}
Hence, given the dissipation rate $\epsilon $ and the dissipation
length-scale $k_d^{-1}$, we can estimate $\nu $.

Here we state another assumption which is used in section 2 of this paper.
We assume that the dissipation length-scale for all the energies, i.e., $%
E^{\pm }(k),E^u(k)$, and $E^b(k)$, are the same and are equal to $k_d^{-1}.$

\newpage\

\end{document}